\documentclass[twocolumn,preprintnumbers,amsmath,amssymb,prl]{revtex4}
\usepackage{graphicx}% Include figure files
\usepackage{bm}% bold math

\begin{document}
\textbf{Kaczorowski \emph{et al.} Reply}: In the preceding Comment \cite{Uhlirova}, Uhlirova \emph{et al.} state that the single crystals of Ce$_2$PdIn$_8$ investigated in Ref. \cite{PRL218} were contaminated by some amount of CeIn$_3$, which resulted in misinterpretation of the observed antiferromagnetic (AF) ordering below 10 K. They demonstrate the result of their energy dispersive X-ray (EDX) analysis that clearly indicates a sandwich-like character of single crystals grown by technique similar to that applied in Ref. \cite{PRL218}, with well defined regions of Ce$_2$PdIn$_8$ and CeIn$_3$. Indeed, our recent metallurgical investigations of the system do corroborate the findings by Uhlirova \emph{et al.} and point out that the presence of CeIn$_3$ in such crystals is hardly avoidable. Because of overlapping of the Bragg peaks due to CeIn$_3$ with those of Ce$_2$PdIn$_8$ and possible residues of indium flux, it is not possible to detect small amount (about 10\%) of the binary phase in X-ray diffraction experiments. Moreover, EDX study on single-crystalline surface is not capable to reveal the presence of a layer of CeIn$_3$ located beneath a relatively thick layer of Ce$_2$PdIn$_8$. These unfortunate shortcomings of the standard sample characterization methods, as well as quite good reproducibility of the results of electrical resistivity and heat capacity measurements performed on a few crystals taken from different batches have led us to incorrect conclusion on the intrinsic character of the antiferromagnetic order in the compound studied. In Fig. 1 we present the low-temperature characteristics of one of those crystals, in which the CeIn$_3$ layer has been removed by polishing the sample down to the thickness of about 100 $\mu$m. Clearly, no phase transition other than that due to superconductivity (SC) is observed, thus definitively ruling out the concept of SC emerging out of AF state.

The authors of the Comment agree with our statement in Ref. \cite{PRL218} that heavy-fermion (HF) superconductivity is a bulk property of Ce$_2$PdIn$_8$, yet they do not present any physical data to illustrate the superconducting behavior in their samples. Furthermore, Uhlirova \emph{et al.} pronounce a large spread in the values of the critical temperature measured for their crystals ($T_{c}$ = 0.45-0.7 K), which is at odds with our own findings. Actually, all the single crystals we studied thus far have been found to superconduct below $T_{c} = 0.70 \pm 0.02$ K (as examples see the data in Fig. 1 and in Ref. \cite{PRL218}). Most recently, the very same $T_c$ has been observed for high-quality polycrystalline samples of Ce$_2$PdIn$_8$ \cite{SSC218}. In all cases, the SC transition has been well-defined in both the electrical resistivity and the heat capacity data, while the parameters describing the superconducting state have had values very similar to those reported in Ref. \cite{PRL218}, which undoubtedly manifest its HF character. Most importantly, in an extended range above $T_c$ the behavior of Ce$_2$PdIn$_8$ distinctly differs from the predictions for a Fermi liquid, namely the resistivity is a linear function of the temperature and the specific heat over temperature ratio diverges with decreasing the temperature (see Fig. 1 and Ref. \cite{SSC218}). These features suggest the presence of antiferromagnetic spin fluctuations and hint at an unconventional character of the superconducting state that possibly emerges in the proximity of a quantum critical point instability, alike in the related HF superconductors Ce$T$In$_5$ and Ce$_2T$In$_8$ with $T$ = Co, Rh and Ir \cite{review}. The latter hypothesis seems supported by recently obtained muon spin rotation spectroscopy and inelastic neutron scattering data \cite{ISIS218}, which clearly evidence magnetically-driven superconducting behavior in Ce$_2$PdIn$_8$, even if long-range AF order is actually absent as an intrinsic property of this compound.

\begin{figure}[!ht]
\includegraphics[width=0.7\columnwidth]{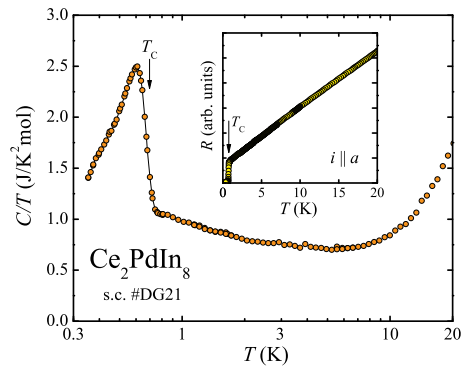}
\caption{(color online) Low-temperature dependencies of the specific heat over temperature ratio and the electrical resistance along the $a$-axis (inset) of single-crystalline Ce$_2$PdIn$_8$.}
\end{figure}

\bigskip

\noindent D. Kaczorowski, A. P. Pikul, D. Gnida, and V. H. Tran
\par Institute of Low Temperature and Structure Research,
\par Polish Academy of Sciences,50-950 Wroc\l aw, Poland
\\

\noindent Received 12 January 2010;\\
PACS numbers: 74.20.Mn;74.70.Tx;74.25.Fy;74.25.Bt

\end{document}